\documentclass[a4paper,11pt]{article}
\pdfoutput=1 

\usepackage{jinstpub} 
\usepackage{subfigure}

\title{\boldmath Effect of plate roughness on the field near RPC plates}

\author[a,c,1]{A. Jash,\note{Corresponding author.}}
\author[a]{N. Majumdar,}
\author[a]{S. Mukhopadhyay}
\author[b]{and S. Chattopadhyay}

\affiliation[a]{Applied Nuclear Physics Division, Saha Institute of Nuclear Physics, Kolkata {700064}, India}
\affiliation[b]{Experimental High Energy Physics Division, Variable Energy Cyclotron Centre, Kolkata {700064}, India}
\affiliation[c]{Experimental High Energy Physics Division, Homi Bhabha National Institute, Mumbai {400085}, India}

\emailAdd{abhik.jash@saha.ac.in}

\abstract{
The inner surfaces of the electrodes encompassing the gas volume of a Resistive Plate Chamber (RPC) 
have been found to exhibit asperities with, grossly, three kinds of features. The desired uniform 
electric field within the gas volume of RPC is expected to be affected due to the presence of these
asperities, which will eventually affect the final response from the detector. In this work, an attempt 
has been made to model the highly complex roughness of the electrode surfaces and compute its 
effect on the electrostatic field within RPC gas chamber. The calculations have been performed 
numerically using Finite Element Method (FEM) and Boundary Element Method (BEM) and the two 
methods have been compared in this context.
}


\arxivnumber{1605.02163} 

\proceeding{13$^{\text{th}}$ WORKSHOP ON RESISTIVE PLATE CHAMBERS AND RELATED DETECTORS,\\
22$-$26 FEBRUARY 2016,\\
GHENT UNIVERSITY, BELGIUM\\
}

\graphicspath{{./../figures/poster_roughness/}}

\begin{document}
\maketitle
\flushbottom

\section{Introduction}
\label{sec:introduction}
Presence of asperities on the inner surfaces of the gas chamber of a Resistive Plate Chamber (RPC) 
is likely to distort the electric field locally  which may give rise to spark, dark current etc. and thus 
lead to gradual degradation of the detector. Glass or Bakelite$^{\tiny \textregistered}$ is commonly 
used as the resistive plates in a RPC and it has been found to develop some asperities on its surface 
in the process of production and handling. The long term operation of the RPCs may also lead to an 
increase in the roughness by affecting the surface through various chemical processes. Early R\&D 
works \cite{saikat-oil_free_bakelite} on Bakelite$^{\tiny \textregistered}$ RPCs for India-based 
Neutrino Observatory (INO) \cite{INO} has shown increase in counting rate and decrease in efficiency 
with time for the RPCs made of Bakelite$^{\tiny \textregistered}$ plates without any surface 
treatment. Performance of the RPCs improved significantly after treating the inner surfaces with 
a suitable fluid.
\\
As the response from an RPC depends critically on the field configuration for a given gas mixture 
and environmental conditions, an extensive study on the effect of roughness on the field 
configuration is deemed necessary. Such a study will help in deciding the acceptable value of 
roughness as well as prediction and interpretation of the experimental data.
In the present work, numerical simulation has been used to estimate this in case of a Bakelite$^{\tiny \textregistered}$-RPC. The RPC has been modelled following the design parameters of a prototype 
fabricated for INO R\&D work. The asperities on the inner surface of the Bakelite$^{\tiny \textregistered}$ electrode have 
been modelled on the basis of the measurement of surface profile of the same Bakelite$^{\tiny \textregistered}$ grades 
used for building the prototype. The calculations have been performed following two numerical methods, 
namely, Finite Element Method (FEM) and Boundary Element Method (BEM). 
\section{Surface Roughness Analysis}
\label{sec:roughnessAnalysis}
\begin{figure}[ht]
 \centering
  \subfigure[]{
  \includegraphics[width=.4\textwidth, height=3.8cm]{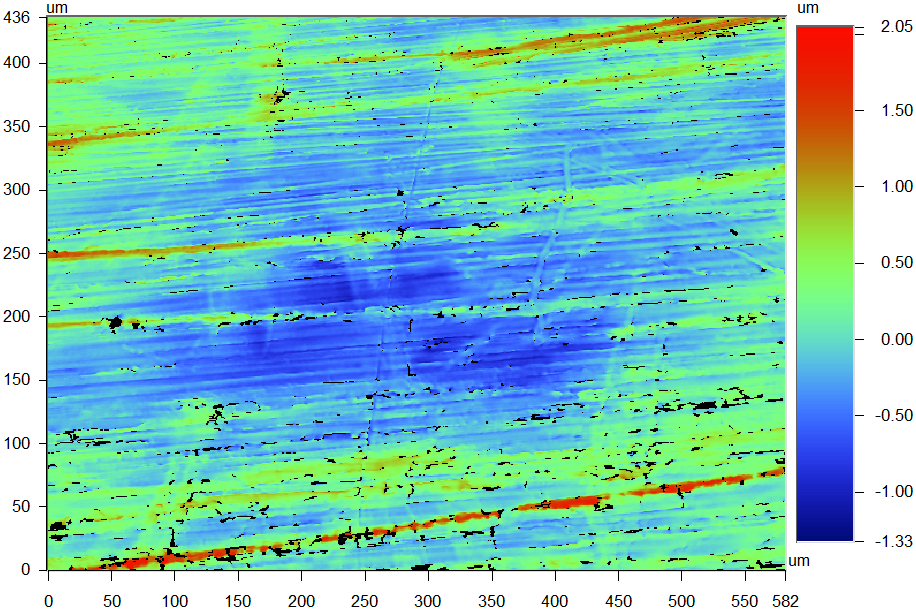}
  \label{fig:roughness_1_cut}
  }
  \hspace{0.3cm}
  \subfigure[]{
  \includegraphics[width=.4\textwidth, height=3.8cm]{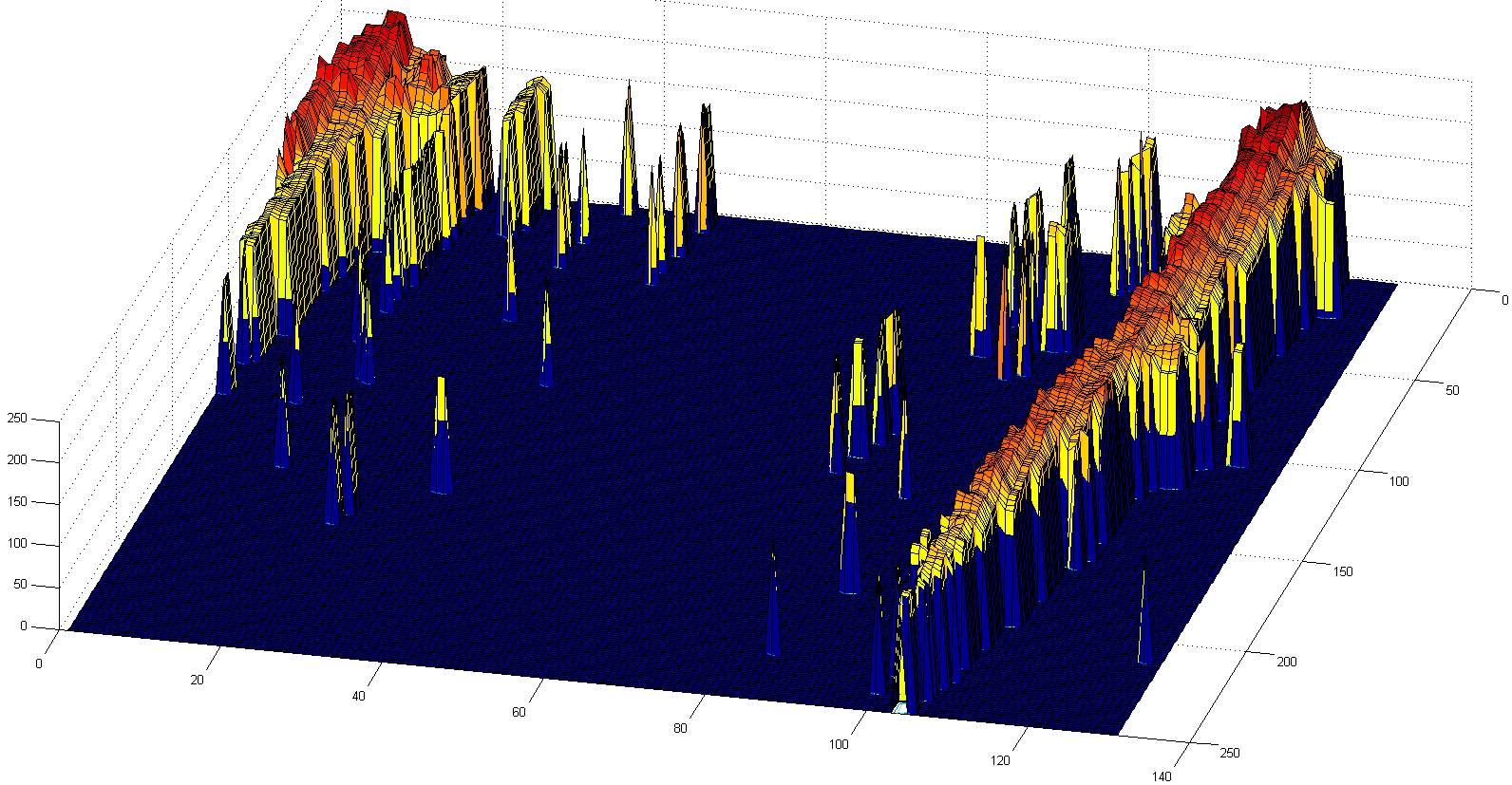}
  \label{fig:roughness_1_3D}
  }
 \caption{\subref{fig:roughness_1_cut} Typical surface morphology of Bakelite$^{\tiny \textregistered}$ P-120 and 
 \subref{fig:roughness_1_3D} the corresponding 3D image after MATLAB analysis.}
\label{fig:roughness_1}
\end{figure}

\begin{figure}[ht]
 \centering
  \subfigure[]{
  \includegraphics[width=.31\textwidth, height=3cm]{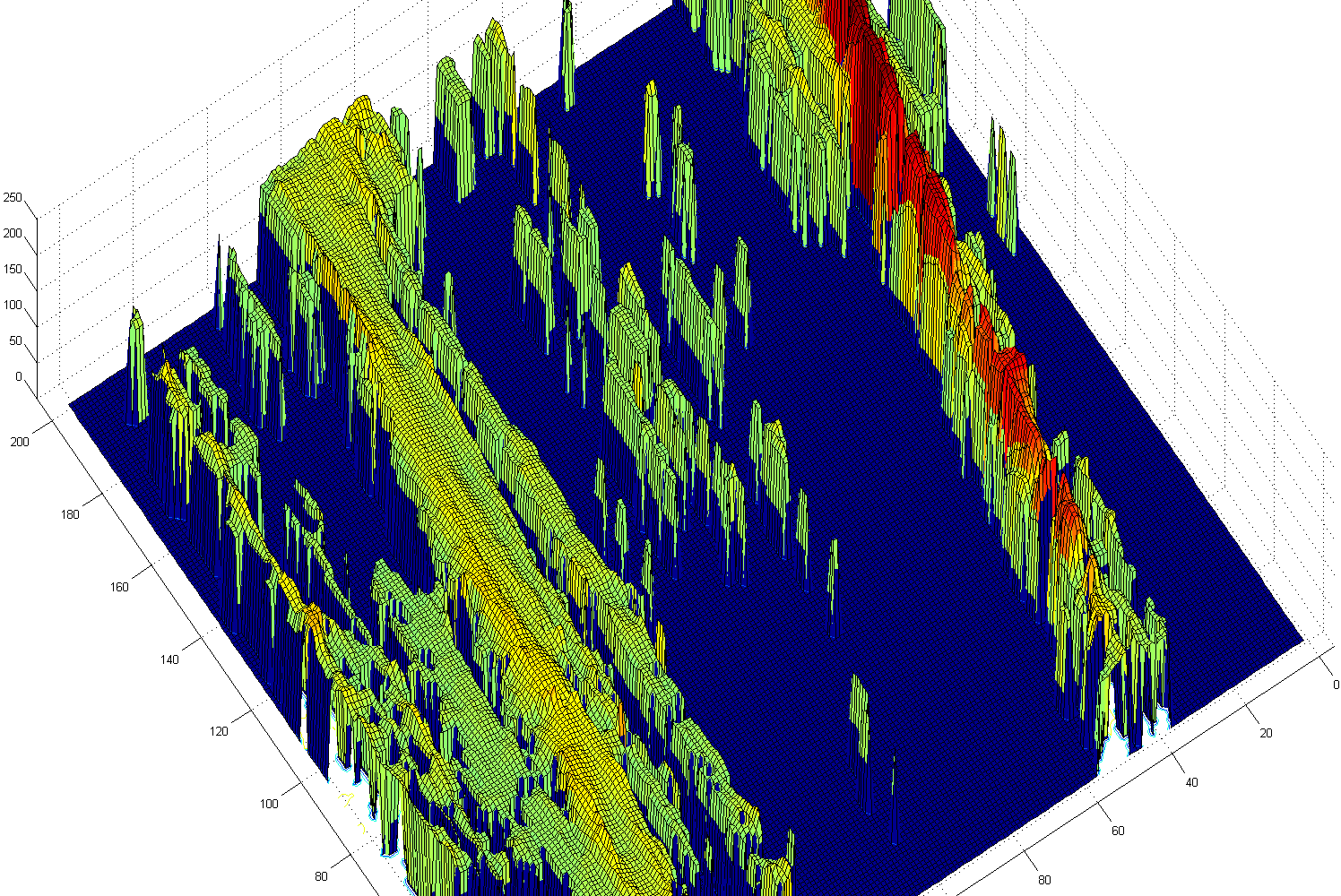}
  \label{fig:roughness_2_3D}
  }
  \subfigure[]{
  \includegraphics[width=.31\textwidth, height=3cm]{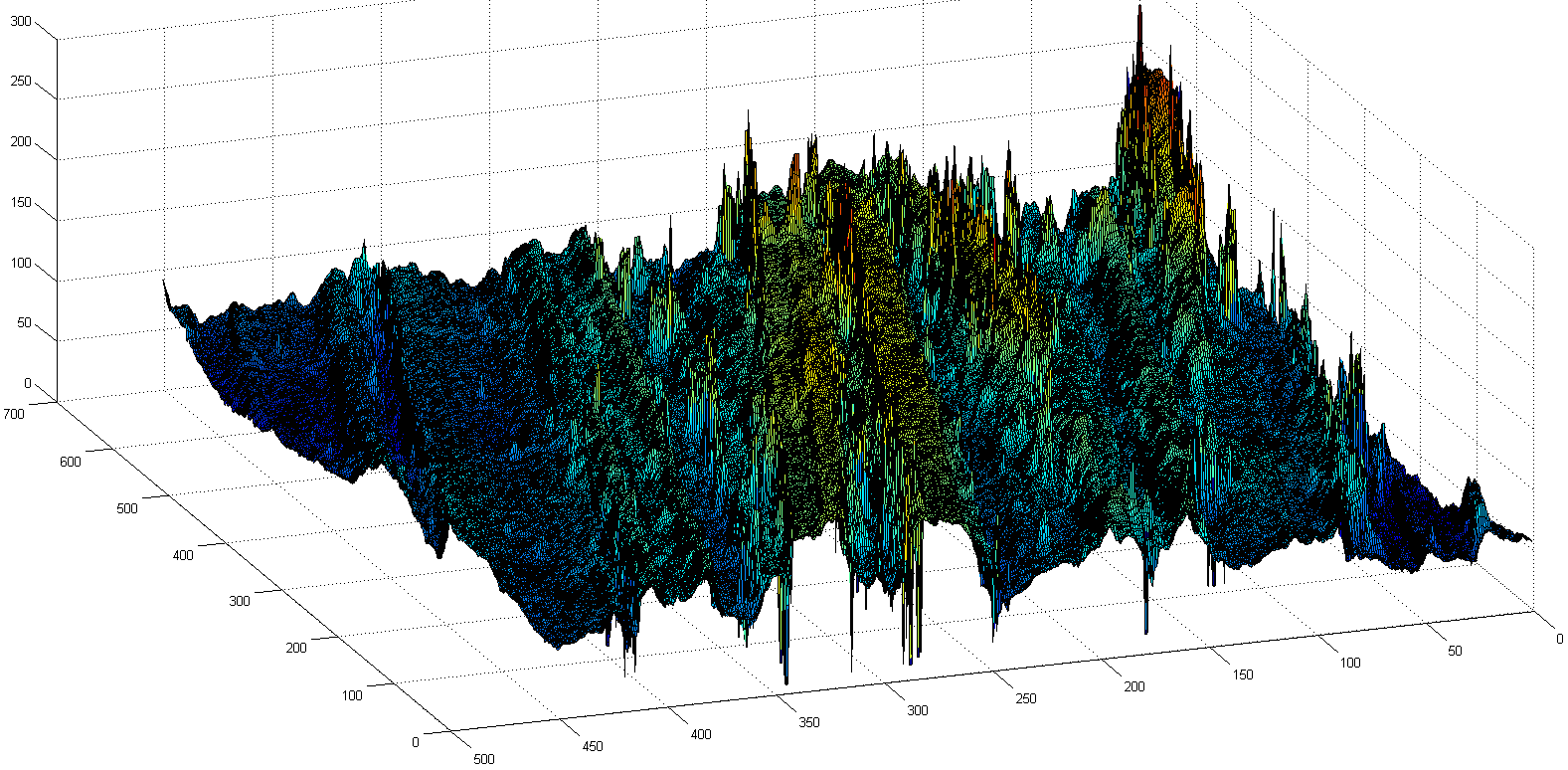}
  \label{fig:roughness_6_3D}
  }
  \subfigure[]{
  \includegraphics[width=.31\textwidth, height=3cm]{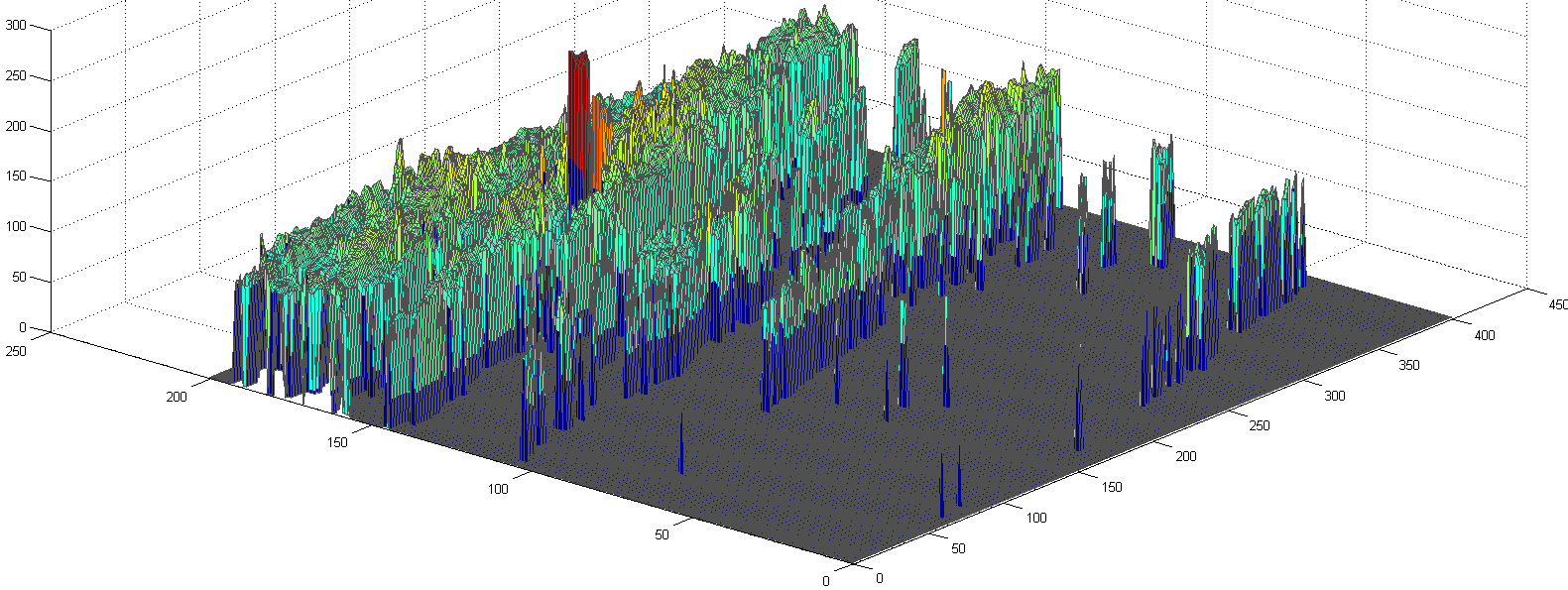}
  \label{fig:roughness_sample2_3_3D}
  }
 \caption{3D images of different parts of Bakelite$^{\tiny \textregistered}$ sample showcasing different building blocks of the 
 surface: \subref{fig:roughness_2_3D} ridges made of boxes and prism-shaped blocks, 
 \subref{fig:roughness_6_3D} spikes sitting on a wavy profile, \subref{fig:roughness_sample2_3_3D}
 spikes of different heights and widths.}
\label{fig:roughness_2_6_3}
\end{figure}

BRUKER ContourGT-K 3D optical microscope has been used to image different samples of a Bakelite$^{\tiny \textregistered}$
plate. Several samples have been made at several locations with scanning area of dimension 
640 $\mu$m $\times$ 480 $\mu$m (shown in figure \ref{fig:roughness_1_cut}, for example). 
The surface images are further analyzed using MATLAB \cite{web_MATLAB}. The dimensions of 
different surface structures have been found out from the color scheme and then 3D images have been 
generated keeping only those structures whose height crosses a certain color threshold value so that 
the building blocks of the surface can be seen clearly. 
The 3D image of figure \ref{fig:roughness_1_cut} after MATLAB analysis has been 
shown in figure \ref{fig:roughness_1_3D}.
The 3D images of few more measurements have been shown in figure \ref{fig:roughness_2_6_3} 
from which the main building blocks of the surface can be found out.
From visual inspection, the following three kind of gross structures in varying dimensions have been 
found : (i) spike, (ii) ridge and (iii) wave.
It has been found out from the data that the average roughness of the surface varies between 150 - 300 nm 
while the range (distance between the peak and the valley) lies between 1 - 5 $\mu$m.
\section{Numerical Modelling of the Surface}
\label{sec:numericalmodelling}
A simplified model of RPC with dimension 5 cm $\times$ 5 cm, consisting of only the Bakelite$^{\tiny \textregistered}$ plates 
and the conductive coats have been modelled to use the computational resources economically, 
with the mentioned asperities distributed on the inner surface of the upper Bakelite$^{\tiny \textregistered}$ plate following 
a definite pattern. A voltage difference of 12 kV has been applied between the coats. 
The three features of the rough surface along with their modelling scheme have been 
discussed below:
\begin{enumerate}
	\item Spike - 
	The randomly distributed spikes have been modelled using boxes of different heights and widths 
	although with a periodic distribution as shown in figure \ref{fig:geometry_boxDistribution}. 
The height and the width of the boxes have been increased individually (along X and Y direction respectively) 
 to a maximum limit of 1 $\mu$m and 4 $\mu$m, respectively, and then reduced again to reach the 
 smallest values along the respective directions.
For the series of boxes, placed along the diagonal direction both the height and width have been 
increased in multiples of the smallest box and reduced back to the smallest values. The reason 
behind the scheme has been to study the individual as well as the combined effect of the variation 
in height and width of the box in a systematic way.
\begin{figure}[ht]
 \centering
  \subfigure[]{
  \includegraphics[width=.34\textwidth, height=3cm]{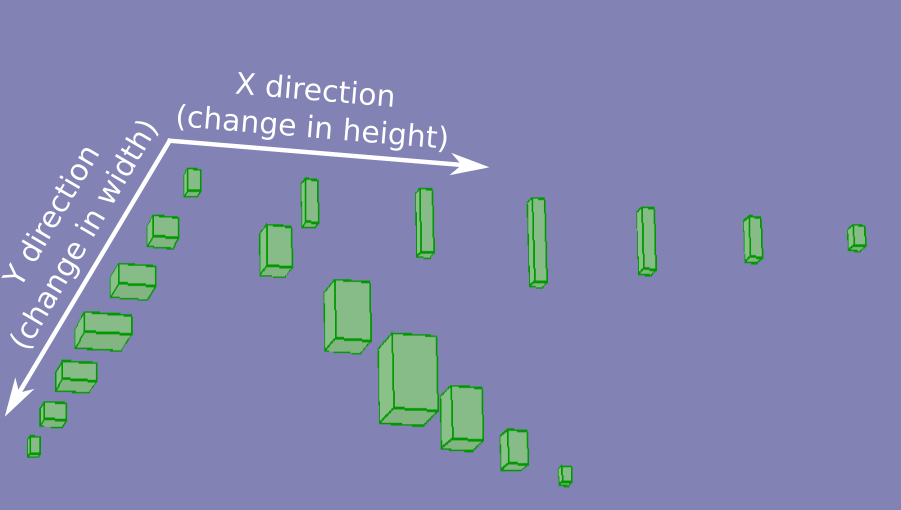}
  \label{fig:geometry_boxDistribution}
  }
  \subfigure[]{
  \includegraphics[width=.29\textwidth, height=3cm]{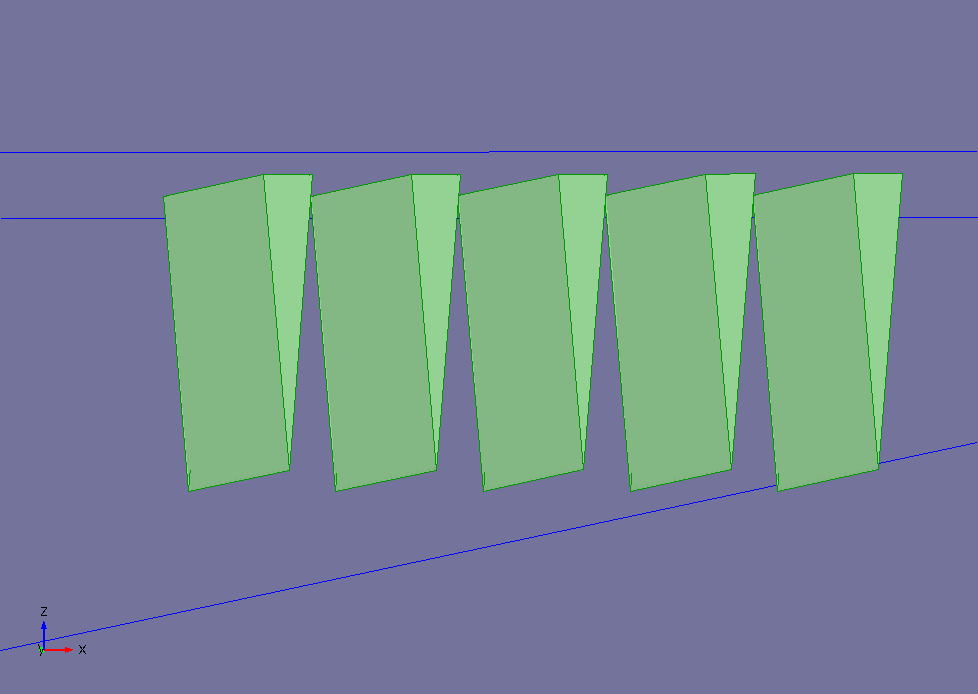}
  \label{fig:geometry_ridges}
  }
  \subfigure[]{
  \includegraphics[width=.29\textwidth, height=3cm]{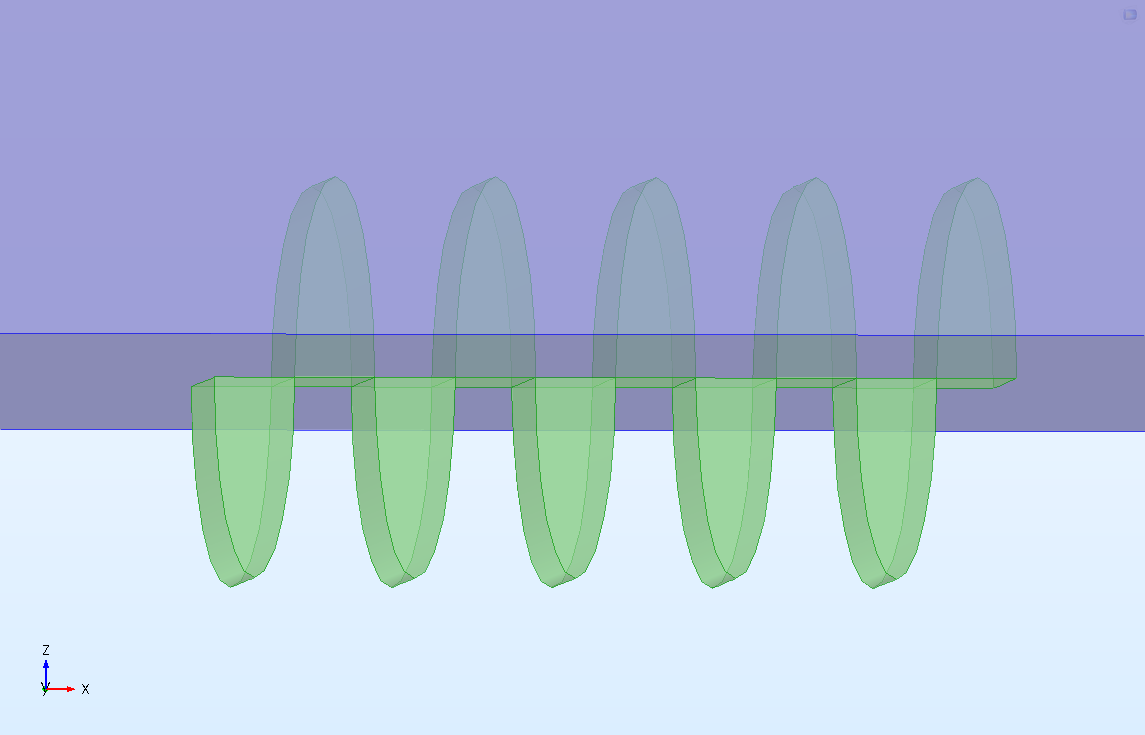}
  \label{fig:geometry_wavy}
  }
 \caption{Surface roughness of \subref{fig:geometry_boxDistribution} spike-like structure modelled 
 using a distribution of boxes of different heights and widths (dimension of smallest box: 
 1 $\mu$m $\times$ 1 $\mu$m $\times$ 250 nm) separated by 8 $\mu$m from each other, 
 \subref{fig:geometry_ridges} ridge-like structures modelled using a series of triangular prism-like 
 blocks each of height 5 $\mu$m, base cross-section 25 $\mu$m $\times$ 650 $\mu$m and a pitch 
 of 50 $\mu$m, \subref{fig:geometry_wavy} wave-like structure modelled using 
 a sinusoidal wave having amplitude of 10 $\mu$m and a wavelength of 40 $\mu$m.}
\label{fig:geometry_box_ridge}
\end{figure}
\item Ridge -
The ridge-like structures have been modelled using five triangular prism-like blocks placed on the 
inner surface of the upper plate with their sharp edges facing the gas volume as shown in figure
\ref{fig:geometry_ridges}. The ridges have been placed parallel to each other along the X-direction.
\item Wave -
The wave-like feature of the surface has been modelled in the shape of a sinusoidal wave as 
shown in figure \ref{fig:geometry_wavy}. The roughness feature has been incorporated on the 
upper Bakelite$^{\tiny \textregistered}$ plate as mentioned earlier with the trenches of depth 10 $\mu$m implanted on 
the plate material.
\end{enumerate}
Two different toolkits, COMSOL Multiphysics$^{\tiny \textregistered}$ \cite{COMSOL} v5.2 based 
on FEM and neBEM \cite{neBEM} v1.8.16 based on BEM have been used to calculate the field.
In COMSOL$^{\tiny \textregistered}$, the whole geometry has been meshed using free tetrahedral 
elements except the conductive coating region which has required a special treatment due to its 
very small thickness \cite{paper-RPC_field}. 
In neBEM, the surfaces of the geometry have been discretized using rectangular or triangular elements 
according to the shape of the geometrical components. The geometry has been built in 
Garfield \cite{web_Garfield} using Constructive Solid Geometry (CSG) approach.
\\
The boxes have been placed 8 $\mu$m away from each other as an optimum value of separation, 
as closer placement requires a finer mesh to resolve the geometry which in turn increases the 
consumption of computational resources. Another difficulty in solving the models arises due to the 
presence of sharp edges, whose effect can be observed only by using infinitesimally small mesh. 
\section{Result}
\label{sec:result}
All the plots have shown the variation in the value of Z-component of the electric field ($E_{z}$) 
for different features of the surface roughness as this is the major component of electric field 
which can affect the RPC response.
\subsection{Effect of the artefacts}
\label{result-artefacts}
\begin{figure}[ht]
  \centering
  \subfigure[]{
  \includegraphics[width=.45\textwidth]{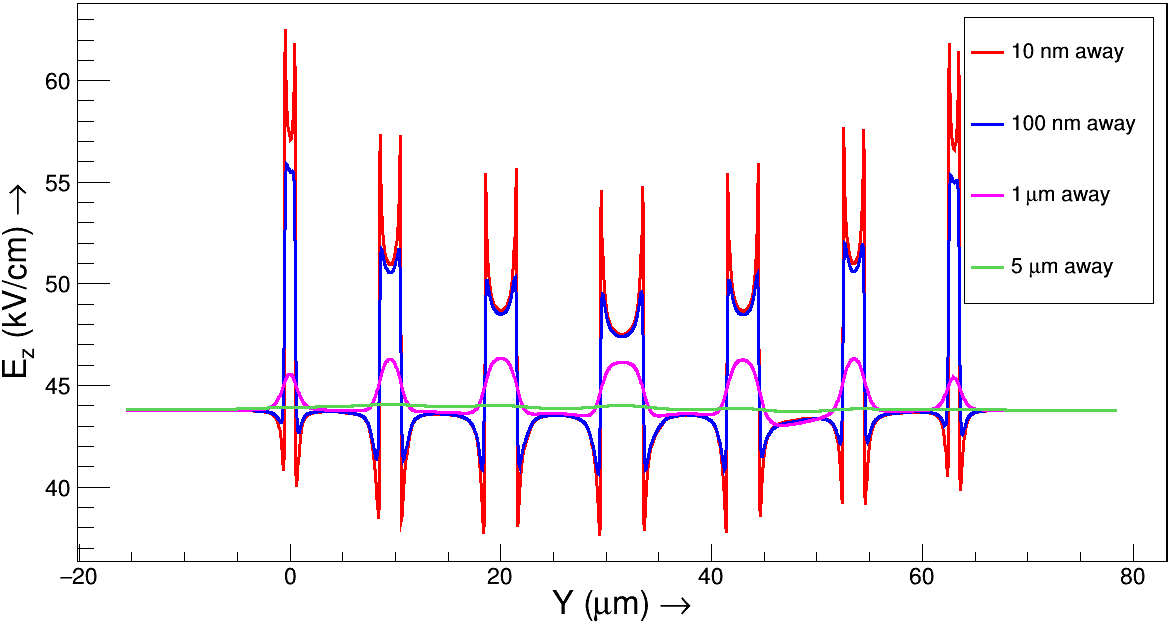}
  \label{fig:box_Ez_Vs_Y}
  }
  \hspace{0.1cm}
  \subfigure[]{
  \includegraphics[width=.45\textwidth]{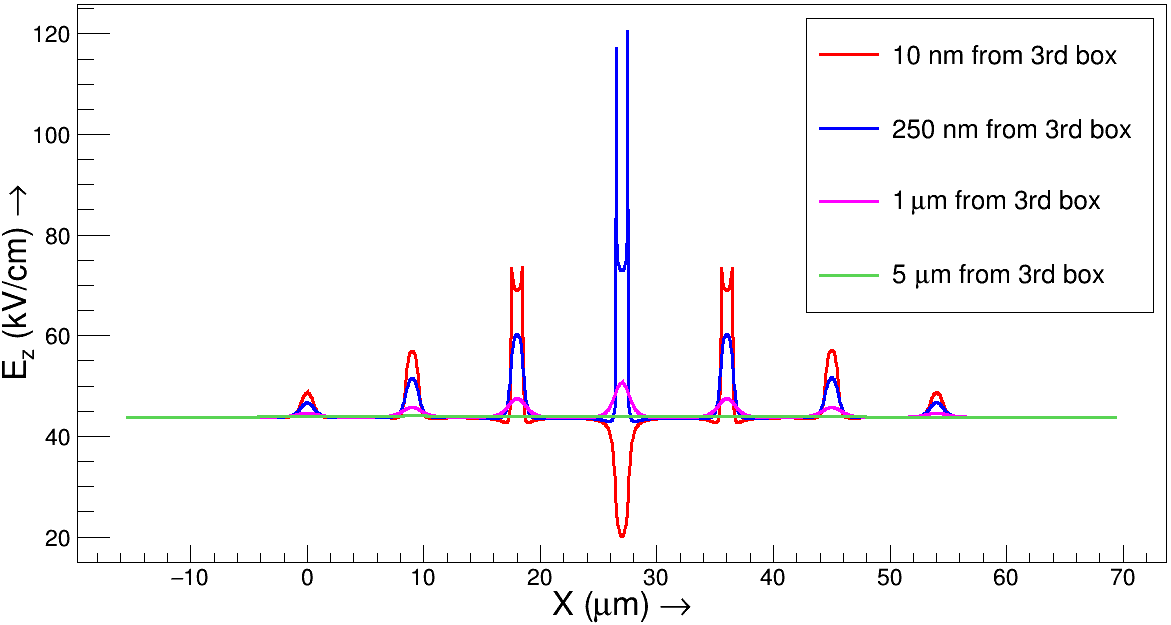}
  \label{fig:box_Ez_Vs_X}
  }
 \caption{Results for the model representing spike-like structures: \subref{fig:box_Ez_Vs_Y} E$_{z}$ vs. Y plot along lines at different heights from the 
tip of the smallest box (affected area increases with the width of the box), \subref{fig:box_Ez_Vs_X} E$_{z}$ 
vs. X plot along lines at different heights from the tip of the third box (height = 750 nm) showing 
the effect of height of the boxes.}
\label{fig:box_Ez_Vs_X_Y}
\end{figure}

The effect of the width and height of the box structures on the nearby electrostatic field, as found 
from neBEM can be seen from the figures \ref{fig:box_Ez_Vs_Y} and \ref{fig:box_Ez_Vs_X} respectively. 
The field is found to increase by 30.5\% maximum at 10 nm away and can rise up to 42.8\% at the 
edges while it falls close to the regular value (within 0.6\%) around 5 $\mu$m.
\begin{figure}[ht]
 \centering
  \subfigure[]{
  \includegraphics[width=.45\textwidth, height=3.8cm]{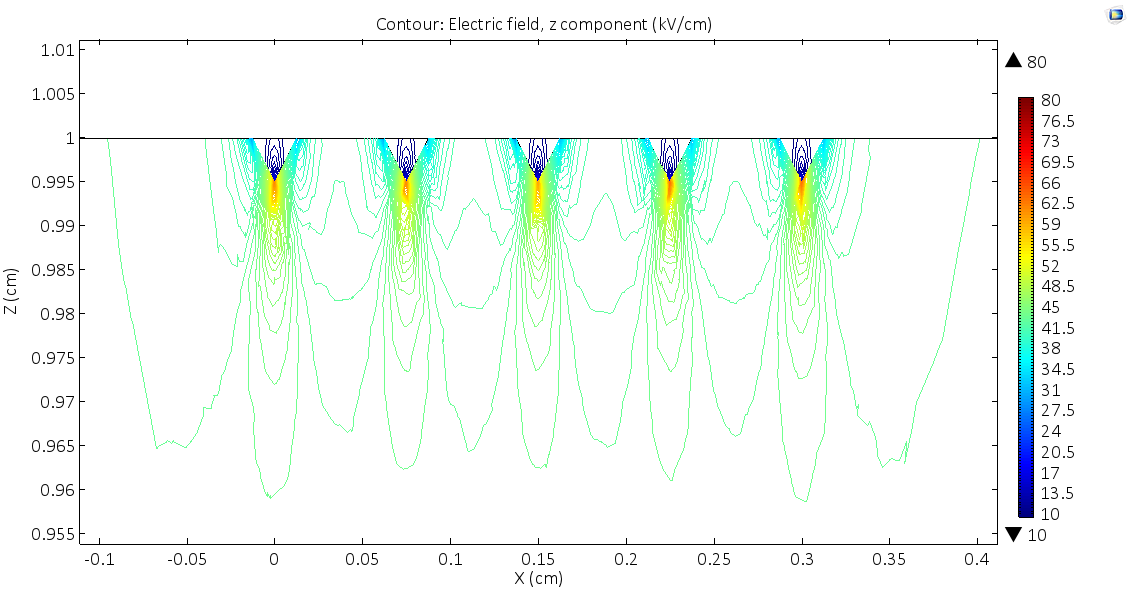}
  \label{fig:ridges_Ez_contour_COMSOL}
  }
  \hspace{0.1cm}
  \subfigure[]{
  \includegraphics[width=.45\textwidth, height=3.8cm]{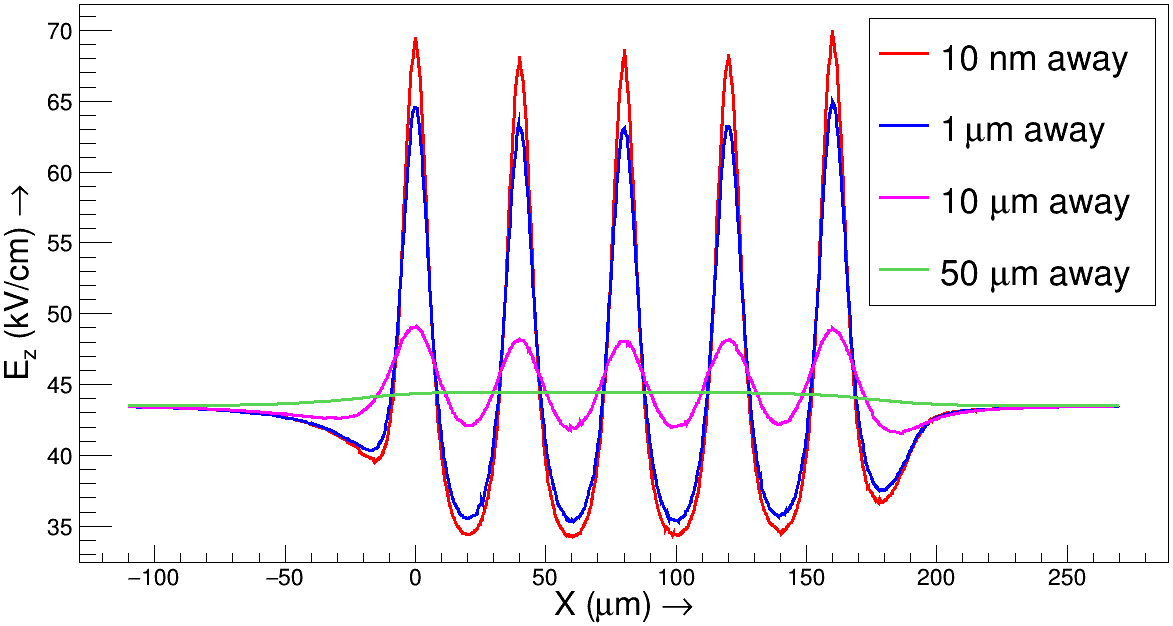}
  \label{fig:wavy_EzVsX_COMSOL}
  }
 \caption{\subref{fig:ridges_Ez_contour_COMSOL} Contour plot of E$_{z}$ near the ridge structure 
 in Y-Z plane from COMSOL$^{\tiny \textregistered}$, \subref{fig:wavy_EzVsX_COMSOL} E$_{z}$ vs. 
 X plot along the lines at different heights from the peaks of the wave shaped profile.}
\label{fig:ridges_result}
\end{figure}
\\
The contour plot of E$_{z}$ in the vicinity of the ridge-like structures can be seen from the figure 
\ref{fig:ridges_Ez_contour_COMSOL} as calculated using COMSOL$^{\tiny \textregistered}$.
The value of E$_{z}$ has been found to deviate from the value in a regular region by 46.2\% to 
0.6\% for a distance 10 nm to 50 $\mu$m away from the structures.
\\
From figure \ref{fig:wavy_EzVsX_COMSOL}, the effect of wavy structure is found up to 50 $\mu$m 
away from the peaks where the relative deviation in the field with respect to its normal 
or regular value reduces to 1.9\% only while it is 59.2\% within 10 nm of the peak and - 21.2\% at 
the troughs.
\subsection{Comparison of FEM and BEM}
\label{result-comparison}
To compare the two solvers, the value of E$_{z}$ has been computed for the spike-like 
structure (see figure \ref{fig:geometry_boxDistribution}) using both the solvers and their 
performance in terms of different parameters has been shown in table \ref{tab:FEMBEM_comparison}.
In case of COMSOL$^{\tiny \textregistered}$, the quoted 
values  are the ones that have been taken by the solver while using the optimized meshing scheme,
which itself may be a time-consuming affair. 
The fifth column of the table shows the convergence criterion for COMSOL$^{\tiny \textregistered}$ 
which is the relative error (R.E.) in the calculated values with respect to the set limit (10$^{-5}$ here)
for the last two iterations. COMSOL$^{\tiny \textregistered}$ has required to iterate 190 times to 
achieve this. For neBEM, the error has been estimated by finding out the deviation in the calculated 
values at the collocation points with respect to the given values as obtained from the boundary 
conditions. The maximum error has been found to be 10$^{-4}$ on one of the elements in this case.
\begin{table}[htbp]
\centering
\caption{\label{tab:FEMBEM_comparison} Comparison of FEM and BEM solvers in solving the model 
of distribution of boxes.}
\smallskip
\begin{tabular}{|c|c|c|c|c|}
\hline
Solver & Elements & Time & Memory & Error/ Convergence \\
\hline
COMSOL$^{\tiny \textregistered}$& 2 $\times$ 10$^{6}$ & 15 mins & $\sim$ 4 GB & 10$^{-5}$ in 190 iterations \\
neBEM & 17 $\times$ 10$^{3}$ & 2 hours & $\sim$ 8.8 GB & R.E.|$_{max}$ = 10$^{-4}$ \\
\hline
\end{tabular}
\end{table}
\section{Summary}
\label{sec:summary}
 The taller and narrower spikes affect the field most, though the affected region stretches with the 
 increase in width.
Depending on the shape and size of the asperities the value of electrostatic field near the asperities 
increases by up to 60\% of the regular value. The distortion in the field is found 
to be confined within a small region (10 - 200 $\mu$m). This distorted field map is likely to affect 
the gas transport properties and production of avalanche close to the resistive plates and alter 
the detector behavior. Considering the overall distribution of asperities on the surface, the detector 
response near the plates may be significantly altered. As a result the effects of surface asperities 
need to be studied in further detail and, depending on the results of the study, surface treatment 
may be planned.

\acknowledgments

We would like to thank Raveendrababu Karanam (IITM) and Dr. Prafulla Behera (IITM) for their help 
in surface roughness measurements and Prof. Sandip Sarkar (SINP) for his help with the MATLAB 
analysis. We would also like to thank the members of INO collaboration for their helpful suggestions 
at different stages of the work and also for the financial support.

\end{document}